\documentclass[a4paper,10pt,twoside,twocolumn]{article}
\usepackage{header-private}
\newcommand{\bu}{\mathbf{u}}
\newcommand{\bg}{\mathbf{g}}

\newcommand{\Pset}{\mathcal{P}}
 
\newcommand{\bnu}{{\bm{\nu}}}
\newcommand{\bs}{{\bm{s}}}

\renewcommand{\xt}{{\mathcal{X}^t}}
\newcommand{\Det}[1]{{\left|#1\right|}}
\newcommand{\mPker}{\mP^\mathrm{he.g}}

\newcommand{\sympkf}{SymPKF }
\newcommand{\code}[1]{\textit{#1}}

\begin{document}

\title{\vspace*{-0.75 cm} SymPKF: a symbolic and computational toolbox for the design of 
parametric Kalman filter dynamics}
\author{O.~Pannekoucke~\footnote{corresponding author: olivier.pannekoucke@meteo.fr}~$^{123}$ and 
Ph.~Arbogast~$^4$\\
$^1$ INPT-ENM, Toulouse, France.\\
$^2$ CNRM, Université de Toulouse, Météo-France, CNRS, Toulouse, France.\\
$^3$ CERFACS, Toulouse, France.\\
$^4$ M\'et\'eo-France, Toulouse, France.
} 
\maketitle

\begin{abstract}	
Recent researches in data assimilation lead to the introduction of the parametric 
Kalman filter (PKF): an implementation of the Kalman filter, where the 
covariance matrices are approximated by a parameterized covariance model. 
In the PKF, the dynamics of the covariance during the forecast step relies on 
the prediction of the covariance parameters. Hence, the design of the parameter 
dynamics is crucial while it can be tedious to do this by hand.
This contribution introduces a python package, \sympkf, able to compute PKF dynamics 
for univariate statistics and when the covariance model is parameterized from the 
variance and the local anisotropy of the correlations. The ability of \sympkf to 
produce the PKF dynamics is shown on a non-linear diffusive advection (Burgers equation) 
over a 1D domain and the linear advection over a 2D domain. The computation of the PKF 
dynamics is performed at a symbolic level, but an automatic code generator is also 
introduced to perform numerical simulations. A final multivariate example 
illustrates the potential of \sympkf to go beyond the univariate case.
\end{abstract}

\vskip 0.3cm
\noindent \textbf{keywords: } Parametric Kalman filter, symbolic computation, 
code generation.

\section{Introduction}

The Kalman filter (KF) \citep{Kalman1960JBE} is one of the backbone of data assimilation. 
This filter represent the dynamics of Gaussian 
distribution all along the analysis and forecast cycles, and takes the form of two equations representing the 
evolution of the mean and of the covariance of the Gaussian distribution.

While the equations of the KF are simple linear algebra, the large dimension of linear space encountered in the 
realm of data assimilation make the KF impossible to handle, and this is particularly true for the forecast step.
This limitation has motivate some approximation of covariance matrix to make the KF possible. For instance, in 
ensemble method \citep{Evensen2009book}, the covariance matrix is approximated by a sample estimation, where the 
time evolution of the covariance matrix is then deduced from the forecast of each individual sample.
In the parametric Kalman filter (PKF) \citep{Pannekoucke2016T,Pannekoucke2018T,Pannekoucke2018NPG}, 
the covariance matrix is approximated by a parametric
covariance model, where the time evolution of the matrix is deduced from the time integration of the parameters' evolution equations. 

One of the major limitation for the PKF is the design of the parameter evolution equations 
While it is not mathematically difficult, this step requires the 
calculus of many terms which is made difficult by hand and might include 
mistake during the computation. To facilitate the derivation of 
the parametric dynamics and certify the correctness of the resulting system a
symbolic derivation of the dynamics would be welcome.

The goal of the package \sympkf\footnote{\url{https://github.com/opannekoucke/sympkf}} 
\citep{Pannekoucke2021Z} is to facilitate the computation of 
the PKF dynamics for a particular class of covariance model, the VLATcov model, 
which is parameterized by the variance and the anisotropy. 
The symbolic computation of the PKF dynamics relies on a 
computer algebra system (CAS) able to handle abstract mathematical expression.  
A preliminary version has been implemented with 
\textrm{Maxima}\footnote{\url{http://maxima.sourceforge.net/}}. However,
in order to create and integrated framework which would include the design 
of the parametric system as well as its numerical evaluation, the  
symbolic python package Sympy \citep{Meurer2017PCS} has been preferred for the 
present implementation. In particular, \sympkf comes with an automatic code generator 
so to provide an end-to-end exploration of the PKF approach, from the computation 
of the PKF dynamics to its numerical integration. 

The paper is organized as follows. The next section provides the background 
on data assimilation and introduces the PKF. The Section~\ref{sec3} focuses on the 
PKF for univariate VLATcov models, in the perspective of its symbolic computation by 
a CAS. Then, the package \sympkf is introduced in Section~\ref{sec4} 
from its use on the non-linear diffusive advection (the Burgers' equation) over a 
1D domain. A numerical example illustrates the use of the automatic code generator provided 
in \sympkf. Then, the example of the linear advection over a 2D domain shows 
the ability of \sympkf to handle 2D and 3D domains. 
The section ends with a simple illustration of a multivariate situation, which also 
show that \sympkf applies on a system of prognostic equations. 
The conclusion is given in Section~\ref{sec5}.

\section{Description of the PKF}\label{sec2}

\subsection{Context of the numerical prediction}\label{sec2.1}

Dynamics encountered in geosciences is given as a system of partial differential equation 
\begin{equation}\label{eq_pde}
	\partial_t \xs = \Mr(t,\partial \xs),
\end{equation}
where $\xs(t,\px)$ is the state of the system and denotes either a scalar 
field or multivariate fields in a coordinate system $\px=(x^i)_{i\in[1,d]}$ 
where $d$ is the dimension the geographical space ; $\partial \xs$ are the 
partial derivatives with respect to the coordinate system at any orders,
with the convention that order zero denotes the field $\xs$ itself ; and 
$\Mr$ denotes the trend of the dynamics.
A spatial discretization (\eg by using finite differences, finite elements,
finite volumes, spectral decomposition, \textit{etc}) transforms \Eq{eq_pde} into
\begin{equation}\label{eq_edo}
\partial_t \xs = \Mr(t,\xs),
\end{equation}
where this time, $\xs(t)$ is a vector, and $\Mr$ denotes this time the discretization of 
the trend in \Eq{eq_pde}.
Thereafter, $\xs$ can be seen either as a collection of continuous fields
with dynamics given by \Eq{eq_pde} or a discrete vector of dynamics \Eq{eq_edo}.

Because of the sparsity and the error of the observations, 
the forecast $\xf$ is only an estimation of the true state $\xt$, which 
is known to within a forecast-error
defined by $e^f = \xf -\xt$. This error is often modelled as an unbiased random 
variable, $\E{e^f}=0$. In the discrete formulation of the dynamics \Eq{eq_edo},
the forecast-error covariance matrix is given by
$\mP^f = \E{e^f(e^f)^\mathrm{T}}$ where the superscript $^\mathrm{T}$ 
denotes the transpose operator. Since this contribution is focused on the 
forecast step, thereafter the upper script $^f$ is removed for the sake of simplicity.

We now detail how the error-covariance matrix evolves during the forecast 
by considering the formalism of the second-order nonlinear Kalman filter.

\subsection{Second-order nonlinear Kalman filter}\label{sec2.11}

A second-order nonlinear Kalman filter (KF2) is a filter that extends the Kalman filter (KF)
to the nonlinear situations where the error-covariance matrix evolves tangent-linearly 
along the trajectory of the mean state and where the dynamics of this mean is governed 
by the fluctuation-mean interacting dynamics \citep{Jazwinski1970book,Cohn1993MWR}. 
Hence, we first state the dynamics of the mean under the fluctuation-mean interaction, 
then the dynamics of the error covariance.
Note that the choice of the following presentation %, which might appears as unusual, 
is motivated by the perspective of using a computer algebra system to perform 
the computation.

\subsubsection{Computation of the fluctuation-mean interaction dynamics}\label{sec2.2.1}

Because of the uncertainty on the initial condition, the state
$\xs$ is modelized as a Markov process $\xs(t,\px,\omega)$ where $\omega$ stands for 
the stochasticity, while $\xs$ evolves by \Eq{eq_pde}. Hence, $\omega$ lies within a 
certain probability space $(\Omega,\Fr,P)$ where $\Fr$ is a $\sigma-$algebra on $\Omega$ 
(a family of subsets of $\Omega$, which contains $\Omega$ and which is stable for the 
complement and the countable union) and $P$ is a probability measure see \eg 
\cite[chap.2]{Oksendal2003book} ; and 
$\xs(t,\px,\cdot):(\Omega,\Fr)\rightarrow(\R^n,\Br_{\R^n})$ is a 
$\Fr-$measurable function where $\Br_{\R^n}$ denotes the 
Borel $\sigma-$algebra on $\R^n$, where the integer $n$ is either the dimension of the 
multivariate field $\xs(t,x)$ or the dimension of its discretized version $\xs(t)$.
%This time the continuous interpretation of the dynamics is essential to obtain the PKF 
%dynamics. 
The connexion between the Markov process and the parameter dynamics is 
obtained using the Reynolds averaging technique.

So to perform the Reynolds averaging of \Eq{eq_pde}, the first step is to replace 
the random field by its Reynolds decomposition 
$\xs(t,\px,\omega) = \E{\xs}(t,\px)+\eta e(t,\px,\omega)$. In this modelling 
of the random state, $\E{\xs}$ is the ensemble average or the mean state; $e$ is an error 
or a fluctuation to the mean, and it is an unbiased random field, $\E{e}=0$. 
Then, \Eq{eq_pde} reads as
\begin{equation}\label{eq_reynolds}
	\partial_t\E{\xs} +\eta\partial_t e = \Mr(t,\partial \E{\xs}+\eta \partial e),
\end{equation}
where $\eta$ is a control of magnitude introduced to facilitate Taylor's expansion
when using a computer algebra system. 
%Now, the Taylor's expansion in $\eta$ is computed.
\begin{subequations}\label{eq_taylor}
	At the second order, the Taylor's expansion in $\eta$ of \Eq{eq_reynolds} reads
	\begin{multline}\label{eq_taylor_2} 
		\partial_t\E{\xs} +\eta\partial_t e = \Mr(t,\partial \E{\xs})+\\
		\eta \Mr'(t,\partial\E{\xs})(\partial e)+\\
			\eta^2 \Mr''(t,\partial \E{\xs})(\partial e\otimes\partial e),
	\end{multline}
	where $\Mr'$ and $\Mr''$ are two linear operators, the former (the later) 
	refers to the tangent-linear model (the hessian), both computed with respect 
	to the mean state $\E{\xs}$. 
	The first order expansion is deduced from \Eq{eq_taylor_2} by setting $\eta^2=0$, 
	which then reads as
	\begin{multline}\label{eq_taylor_1} 
	\partial_t\E{\xs} +\eta\partial_t e = \Mr(t,\partial\E{\xs})+\\	
		\eta \Mr'(t,\partial \E{\xs})(\partial e).
	\end{multline}
\end{subequations}

By setting $\eta$ to one, the dynamics of the ensemble average is calculated at 
the second order from the expectation of \Eq{eq_taylor_2} that reads as
\begin{multline}\label{eq_fluc_mean}
	\partial_t\E{\xs} = \Mr(t,\partial\E{\xs})+	\\	
	\Mr''(t,\partial\E{\xs})(\E{\partial e\otimes\partial e}),
\end{multline}
where $\partial e\otimes\partial e$ denotes the tensorial product partial derivatives with respect to 
the spatial coordinates, \ie terms as $\partial^k e\partial^m e$ for any positive integers $(k,m)$. Here, we have 
used that the partial derivative commutes with the expectation, $\E{\partial e} = \partial\E{e}$, 
and that $\E{e}=0$.
Because the expectation is a projector, $\E{\E{\cdot}}=\E{\cdot}$, expectation of 
$\Mr(t,\partial\E{\xs})$ is itself.
The second term of the right hand side makes appear the retro action of the error 
onto the ensemble averaged dynamics. Hence, \Eq{eq_fluc_mean} gives the dynamics of the 
error-mean interaction (or fluctuation-mean interaction).

Note that, the tangent-linear dynamics along the ensemble averaged dynamics \Eq{eq_fluc_mean} 
is obtained as the difference between the first order Taylor's 
expansion \Eq{eq_taylor_1} by its expectation, and reads as
\begin{equation}\label{eq_tl}
	\partial_t e = \Mr'(t,\partial\E{\xs})(\partial e).
\end{equation}

Now it is possible to detail the dynamics of the error covariance from the 
dynamics of the error which tangent-linearly evolves along the mean state $\E{\xs}$.

\subsubsection{Computation of the error-covariance dynamics}

In the discretized form, the dynamics of the error \Eq{eq_tl} reads as 
\begin{equation}\label{eq_edotl}
	\frac{de}{dt} = \mM e,
\end{equation}
where $\mM$ stands for the tangent-linear model $\Mr'(t,\partial\E{\xs})$ evaluated 
at the mean state $\E{\xs}$.
So the dynamics of the error-covariance matrix, $\mP=\E{e e^\mathrm{T}}$, is given by 
\begin{subequations}\label{eq_ODEKF}
	\begin{equation}\label{eq_ODEKFa}
		\frac{d\mP}{dt} = \mM\mP + \mP\mM^\mathrm{T}
	\end{equation}
	($\mM^\mathrm{T}$ is the adjoint of $\mM$),
	or its integrated version 
	\begin{equation}\label{eq_ODEKFb}
		\mP(t)=\mM_{t\leftarrow 0}\mP_0\left(\mM_{t\leftarrow 0}\right)^\mathrm{T}
	\end{equation}			
\end{subequations}
where $\mM_{t\leftarrow 0}$ is the propagator 
associated to the time integration of \Eq{eq_edotl}, initiated from the covariance $\mP_0$.

\subsubsection{Setting of the KF2}

Gathering the dynamics of the ensemble mean given by the fluctuation-mean interaction 
\Eq{eq_fluc_mean} and the covariance dynamics \Eq{eq_ODEKF} leads to the second-order closure 
approximation of the extended KF, that is the forecast step 
equations of the KF2.

Similarly to the KF, the principal limitation of the KF2 is the numerical 
cost associated with the covariance dynamics \Eq{eq_ODEKF}:
living in a discrete world, the numerical cost of \Eq{eq_ODEKF} 
dramatically increases with the size of the problem.
As an example, for the dynamics of simple scalar field discretized 
with $n$ grid points, the dimension of its vector representation is $n$, 
while the size of the error-covariance matrix scales as 
$n^2$ ; leading to a numerical cost of \Eq{eq_ODEKF} between $\Or(n^2)$ and $\Or(n^3)$.

We now introduce the parametric approximation of covariance 
matrices which aims to reduce the cost of the covariance dynamics \Eq{eq_ODEKF}.

\subsection{Formulation of the PKF prediction}\label{sec2.2}

The parametric formulation of covariance evolution stands as follows.
If $\mP(\Pset)$ denotes a covariance model featured by 
a set of parameters $\Pset=(p_i)_{i\in I}$, then there exists a set 
$\Pset_t^f$ featuring the forecast  error covariance 
matrix so that $\mP(\Pset_t^f)\approx \mP^f_t$.

Hence, starting from the initial condition 
$\Pset^f = \Pset_0^f$, if the dynamics of the parameters $\Pset^f_t$ is known, 
then it is possible to approximately determine $\mP^f_t\approx\mP(\Pset^f_t)$ 
without solving \Eq{eq_ODEKF} explicitly. 
This approach constitutes the so-called parametric 
Kalman filter (PKF) approximation, introduced by
\cite{Pannekoucke2016T,Pannekoucke2018NPG} (P16, P18).

We now focus on the PKF applied to a particular family of covariance models.

\section{PKF for VLATcov models}\label{sec3}

This part introduces a particular family of covariance models, parameterized by
the fields of variances and of local anisotropy tensor: the VLATcov  models 
\citep{Pannekoucke2020X}.
What makes this covariance model interesting is that its parameters are related
to the error field and thus, it is possible to determine the dynamics of the parameters.
So to introduce VLATcov models, we first present the diagnosis of the variance 
and of the local anosotropy tensor, then we present two examples of VLATcov models and 
we end the section by the description of the dynamics of the parameters.

\subsection{Definition of the fields of variance and of local anisotropy tensor}

From now, we will focus on the forecast-error statistics, so the upperscript $^f$ is
removed for the sake of simplicity. Moreover, for a function $f$, when there 
is no confusion, the value of $f$ at a point $\px$ is written either as $f(\px)$ or 
as $f_\px$.

The forecast error being unbiased, $\E{e}=0$, its variance at a point $\px$ is defined as
\begin{equation}\label{eq_variance}
V(\px) = \E{e(\px)^2}.
\end{equation}
When the error is a random differentiable field, the anisotropy of the two-points correlation function 
$\rho(\px,\py) = \frac{1}{\sqrt{V_\px V_\py}} \E{e(\px)e(\py)}$
is featured, from the second order expansion
\begin{equation}
\rho(\px,\px+\delta\px) \approx 1 - \frac{1}{2}||\delta\px||^2_{\bg_\px},
\end{equation}
by the local metric tensor $\bg(\px)$, and defined as
\begin{equation}\label{eq_g}
\bg(\px) = -\nabla\nabla^\mathrm{T}\rho_\px,
\end{equation}
where $\rho_\px(\py)=\rho(\px,\py)$ \eg 
$$g_{ij}(\px)=-\left(\partial_{y^iy^j}^2 \rho_\px(\py)
\right)_{\py=\px}.$$
The metric tensor is a symmetric positive definite matrix, and it is a 
$2\times 2$ ($3\times 3$) matrix in a 2D (3D) domain.

Note that it is useful to introduce the local aspect tensor \citep{Purser2003MWRa}
whose the geometry goes as the correlation, and defined as the inverse of the metric tensor
\begin{equation}\label{Eq:metric-aspect}
\bs(\px) = \bg(\px)^{-1},
\end{equation}
where the superscript $^{-1}$ denotes the matrix inverse.

What makes the metric tensor attractive, either at a theoretical or at a practical level, 
is that it is closely related to the normalized error 
$\eps = \frac{e}{\sqrt{V}} $ by 
\begin{equation}\label{eq_bgij}
\bg_{ij}(\px) = \E{
	(\pde_{x^i}\eps)
	(\pde_{x^j}\eps)
	}
\end{equation} 
(see \eg \citep{Pannekoucke2020X} for details).

Hence, a VLATcov model is a covariance model characterized by the variance field,
and by the anisotropy field, the latter being defined either by the metric-tensor 
field $\bg$ or by the aspect-tensor field $\bs$. 
To put some flesh on the bones, two examples of VLATcov models are now presented.

\subsection{Examples of VLATcov models}

The covariance model based on the the heterogeneous diffusion operator of 
\cite{Weaver2001QJRMS} is often introduced in numerical weather or ocean prediction
to model heterogeneous correlation functions.
This model own the property that, under the local homogenous assumption
(that is when the spatial derivatives are negligible) 
the local aspect tensors of the correlation functions are twice the local diffusion tensors.
\citep{Pannekoucke2008QJRMSa,Mirouze2010QJRMS}. 
Hence, by defining the local diffusion tensors as half the local aspect tensors, 
the covariance model based on the heterogeneous diffusion equation is a VLATcov model.

Another example of heterogeneous covariance model is the 
heterogenous Gaussian covariance model
\begin{multline}\label{eq_Phegauss}
\mPker(V,\bnu)(\px,\py)= \\
\sqrt{V(\px)V(\py)}
\frac{\Det{\bnu_\px}^{1/4}\Det{\bnu_\py}^{1/4}}{\Det{\frac{1}{2}(\bnu_\px+\bnu_\py)}^{1/2}
}\\
\exp\left(-||\px-\py||^2_{(\bnu_\px+\bnu_\py)^{-1}}\right),
\end{multline}
where $\bnu$ is a field of symmetric positive definite matrices, and 
$|\bnu|$ denotes the matrix determinant.
$\mPker(V,\bnu)$ is a particular case of the class of covariance models
deduced from Theorem~1 of  \cite{Paciorek2004ANIPS}. Again, this covariance 
model own the property that, under local homogeneous assumptions, 
the local aspect tensor is approximately given by $\bnu$, \ie for any point $\px$,
\begin{equation}
	\bs_x\approx\bnu_x.
\end{equation}
Hence, as for the covariance model based on the diffusion equation, 
by defining the field $\bnu$ as the aspect tensor field, the heterogeneous 
Gaussian covariance model is a VLATcov model \citep{Pannekoucke2020X}.

At this stage, all the pieces of the puzzle are put together to build the PKF dynamics.
We have covariance models parameterized from the variance and the local anisotropy,
which are both related to the error field: knowing the dynamics of the error leads 
to the dynamics of the VLATcov parameters. This is now detailed.

\subsection{PKF prediction step for VLATcov models}

When the dynamics of the error $e$ is well approximated from the 
tangent-linear evolution \Eq{eq_tl}, the connection between the covariance 
parameters and the error, \Eq{eq_variance} and \Eq{eq_bgij}, 
makes possible to establish the prediction step of the PKF 
\citep{Pannekoucke2018NPG}, which reads as 
\begin{subequations}\label{eq_pkf}
the dynamics of the ensemble average (at the second-order closure)
\begin{multline}
	\label{eq_pkf_aa}	
	\partial_t\E{\xs} = \Mr(t,\partial\E{\xs})+\\		
		\Mr''(t,\partial\E{\xs})(\E{\partial e\otimes\partial e}),
\end{multline}
coupled with the dynamics of the variance and the metric
\begin{align}
\label{eq_pkf_a} & \pdt V(t,\px) = 2 \E{e\pdt e},\\
\label{eq_pkf_b} & \pdt \bg_{ij}(t,\px) =
			\E{\pdt\left(
			(\pde_{x^i}\eps)
			(\pde_{x^j}\eps)
			\right)
			},
\end{align}
\end{subequations}
where it remains to replace the dynamics of the error (and its normalized version 
$\eps=e/\sqrt{V}$)
from \Eq{eq_tl}, and where property that the expectation operator and the temporal 
derivative commutes, $\pdt\E{\cdot} = \E{\pdt\cdot}$,  
has been used to obtain \Eq{eq_pkf_a} and \Eq{eq_pkf_b}.

The set of equations (\ref{eq_pkf}) is at the heart of the 
numerical sobriety of the parametric approach. 
In contrast to the matrix dynamics of the KF, the PKF approach is designed for the
continuous world, leading to PDEs for the parameters dynamics
in place of ODEs \Eq{eq_ODEKF} for the full matrix dynamics.
For the above mentioned scalar fields, introduced is the computation of the algorithmic 
complexity in section~\ref{sec2.1}, the cost of \Eq{eq_pkf} is $\Or(n)$.
Moreover, the dynamics of the parameters sheds light on the nature 
of the processes governing the dynamics of covariances ; 
and it does not require any adjoint of the dynamics
\citep{Pannekoucke2016T,Pannekoucke2018NPG}.

Note that \Eq{eq_pkf} can be formulated in terms of 
aspect tensors, thanks to the definition \Eq{Eq:metric-aspect}:
since, $\bs \bg = \mI$, its 
time derivative $(\pdt\bnu)\bg+\bnu(\pdt\bg)=0$ leads to the dynamics 
$\pdt \bs = -\bg^{-1}(\pdt\bg)\bs$, and then
\begin{equation}\label{eq_in_aspect}
	\pdt \bs = -\bs(\pdt\bg)\bs,
\end{equation}
where it remains to replace occurrences of $\bg$ by $\bs^{-1}$ in the resulting 
dynamics of the mean, the variance and the aspect tensor.

Hence, the PKF forecast step for VLATcov model is given by either the system
\Eq{eq_pkf} (in metric), or by its aspect tensor formulation thanks to 
\Eq{eq_in_aspect}.
Whatever the formulation considered, it is possible to carry out the calculations using a
formal calculation language. However, even for simple physical processes, the number of
terms in formal expressions can become very large, \eg it is common to have to manipulate
expressions with more than a hundred terms.
Thus, any strategy that simplifies the assessment of PKF systems in advance can quickly become a significant advantage. 

In the following section, we present the splitting method that allows the PKF dynamics to be expressed by bringing together the dynamics of each of the physical processes, calculated individually.

\subsection{The splitting strategy}

When there are several processes in the dynamics \Eq{eq_pde},
the calculation of the parametric dynamics can be tedious even when 
using a computer algebra system. To better use digital resources, a splitting strategy 
can be introduced \citep{Pannekoucke2016T,Pannekoucke2018NPG}. 

While the theoretical background is provided by 
the Lie-Trotter formula for Lie derivatives, the well-known idea of time-splitting is easily catched from a first order Taylor expansion of an Euler numerical scheme:

The computation of a dynamics 
\begin{equation}\label{eq_mul_proc}
	\pdt\xs = f_1(\xs)+f_2(\xs),
\end{equation}
 over a single time step $\delta t$,
 can be done in two times following the numerical scheme
 \begin{equation}
	\left\{
		\begin{array}{l}
			\xs^\star = \xs(t)+\delta t f_1(\xs(t)),\\
			\xs(t+\delta t) = \xs^\star+\delta t f_2(\xs^\star).
		\end{array}
	\right.
\end{equation}
where at order $\delta t$, this scheme is equivalent to 
$\xs(t+\delta t) =\xs(t) +\delta t\left(f_1(\xs(t))+f_2(\xs(t))\right)$, that is the Euler step of 
\Eq{eq_mul_proc}. Because $f_1$ and $f_2$ can be viewed as vector fields,
the fractional scheme, joining the starting point (at $t$) to the end point (at $t+\delta t$), 
remains to going through the parallelogram, formed by the sum of the two vectors, along its sides. 
Since there are two paths joining the extreme points, starting the computation by $f_2$ 
is equivalent to starting by $f_1$ (at order $\delta t$), this corresponds to the commutativity 
of the diagram formed by the parallelogram. 

Appendix~\ref{secA} shows that a dynamics given by \Eq{eq_mul_proc} implies a dynamics of the error, the variance, the metric 
and the aspect, written as a sum of trends. Hence, it is possible to apply a splitting for all these dynamics.

As a consequence for the calculation of the parametric dynamics: calculating the parametric dynamics of
\Eq{eq_mul_proc} is equivalent to calculating separately the parametric dynamics of $\pdt \xs = f_1(\xs)$ 
and $\pdt \xs = f_2(\xs)$, then bringing together the two parametric dynamics into a single
one by summing the trends for the mean, the variance, the metric or the aspect dynamics.
This splitting applies when there are more than two processes and 
appears as a general method to reduce the complexity of the calculation.

\subsection{Discussion/intermediate conclusion}

However, while the computation of the system \Eq{eq_pkf} is straightforward,
since it is similar to the computation of the Reynolds equations
\citep{Pannekoucke2018NPG}, it is painful because of the numerous terms 
it implies, and there is a risk to introduce errors during the computation by hand.

Then, once the dynamics of the parameters is established, it remains to design a numerical 
code to test whether the uncertainty is effectively well represented by the PKF 
dynamics. Again, the design of a numerical code is not necessarily difficult 
but with the numerous terms, the risk of introducing an error is important.

To facilitate the design of the PKF dynamics as well as the numerical evaluation, the 
package \sympkf 
 has been introduced to perform the VLATcov parameter dynamics and 
to generate a numerical code used for the investigations \citep{Pannekoucke2021Z}. 
The next section introduces and details this tool.

\section{Symbolic computation of the PKF for VLATcov}\label{sec4}

In order to introduce the symbolic computation of the PKF for VLATcov model, 
we consider an example: the diffusive non-linear advection, 
the Burgers equation, which reads as
\begin{equation}\label{eq_burgers}
	\pdt u + u \pdx u =\kappa\pdx^2 u,
\end{equation}
where $u$ stands for the velocity field and corresponds to a function of the time 
$t$ and the space of coordinate $x$,
and where $\kappa$ is a diffusion coefficient (constant here).
This example illustrates the workflow leading to the PKF dynamics. 
It consists in defining the system of equation in Sympy, then to compute the 
dynamics \Eq{eq_pkf}, we now detail these two steps.

\subsection{Definition of the dynamics}

\begin{figure}
	\includegraphics[width=7.5cm]{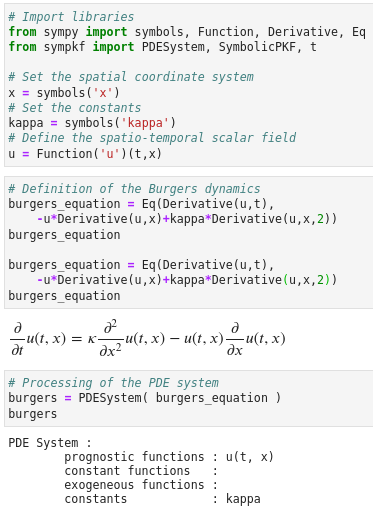}
	\caption{Sample of code and Jupyter notebook outputs for the definition of the Burgers dynamics using
	 \code{sympkf}.
	}\label{fig_burgers_setup}
\end{figure}

The definition of the dynamics relies on the formalism of sympy as shown in 
\Fig{fig_burgers_setup}. The coordinate system is first defined as instances of the 
class \texttt{Symbols}. Note that the time is defined as \texttt{sympkf.t} while
the spatial coordinate is let to the choice of the user, here $x$. 
Then, the function $u$ is defined as an instance of the class \texttt{Function}, as a 
function of $(t,x)$.

In this example, the dynamics consists in a single equation defined as an instance of 
the class \texttt{Eq}, but in the general situation where the dynamics is 
given as a system of equation, the dynamics has to be represented as a python list 
of equations.

A preprocessing of the dynamics is then performed to determine several important quantities
to handle the dynamics: the prognostic fields
(functions for which a time derivative is present), the diagnostic fields (functions
for which there is no time derivative in the dynamics), the constant functions (functions
that only depend on the spatial coordinates), and the constants (pure scalar terms, that
are not function of any coordinate). This preprocessing is performed when 
transforming the dynamics as an instance of the class \texttt{PDESystem}, 
and whose default string output delivers a summary of the dynamics: 
for the Burgers' equation, there is only one prognostic function, $u(t,x)$, 
and one constant, $\kappa$.

The prognostic quantities being known, it is then possible to perform the 
computation of the PKF dynamics, as discussed now.

\subsection{Computation of the VLATcov PKF dynamics}

\begin{figure}
	\includegraphics[width=8cm]{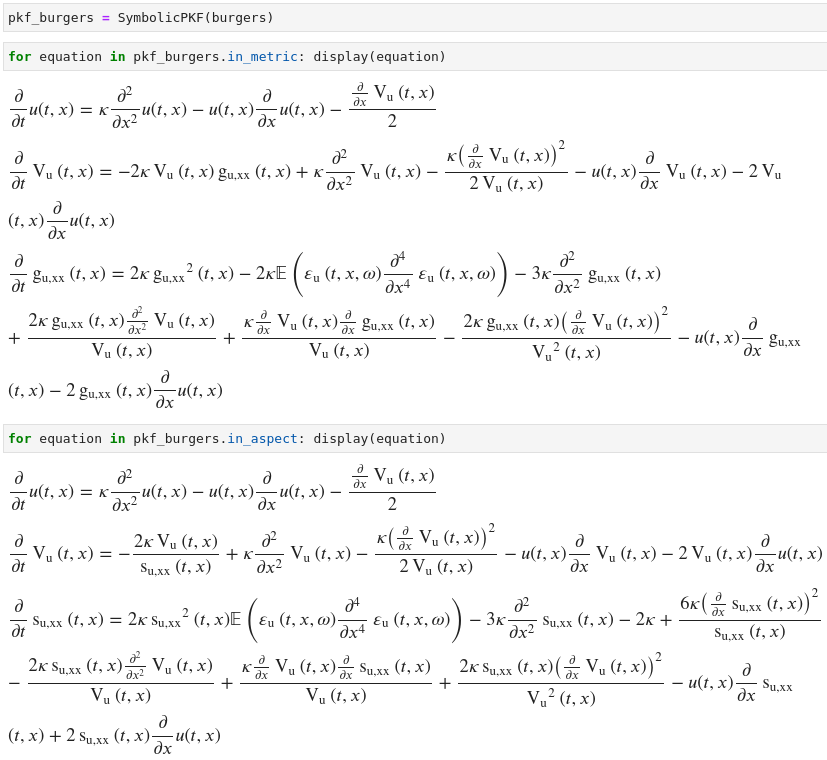}
	\caption{Sample of code and Jupyter notebook outputs: 
	systems of partial differential equations given in metric and in aspect forms, 
	produced by \code{sympkf} when applied to the Burgers' equation \Eq{eq_burgers}.}
	\label{fig_burgers_unclosed}
\end{figure}

\begin{figure}
	\includegraphics[width=8cm]{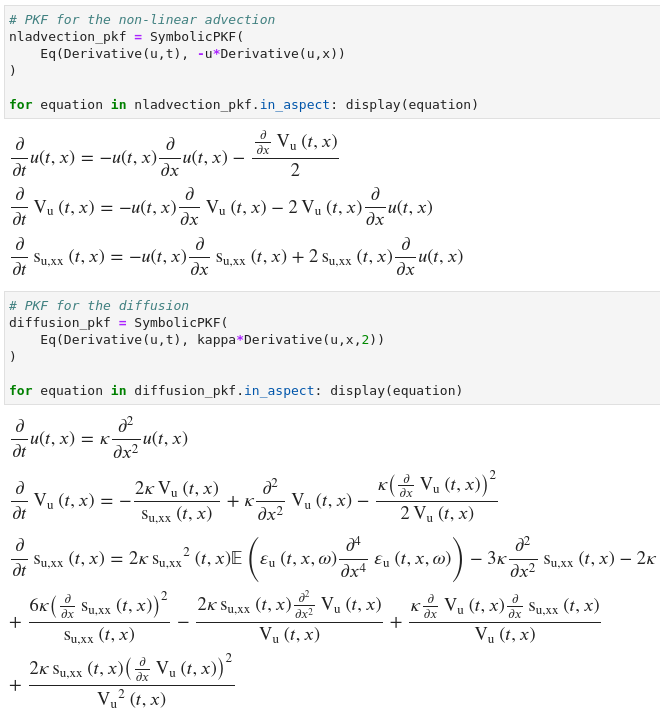}
	\caption{Illustration of the splitting strategy which can be used to compute the PKF
	dynamics and applied here for the Burgers' equation: PKF dynamics of 
	the Burgers' equation can be obtained from the PKF dynamics of the advection (first cell) 
	and of the diffusion (second cell).}
	\label{fig_burgers_splitting}
\end{figure}

Thanks to the preprocessing, we are able to determine what are the VLATcov parameters 
needed to compute the PKF dynamics, that is the variance and the anisotropy tensor 
associated to the prognostic fields. For the Burgers' equation, the VLATcov parameters
are the variance $V_u$ and the metric tensor $\bg_u = (g_{u,xx})$ or its associated 
aspect tensor $\bs_u = (s_{u,xx})$. Note that, in \texttt{sympkf}, the VLATcov parameters 
are labeled by there corresponding prognostic fields so to facilitate their identification.
This labelling is achieved when the dynamics is transformed as an instance of the class 
\texttt{SymbolicPKF}. This class is at the core of the computation of the PKF dynamics from 
\Eq{eq_pkf}. 

% intro of stochasticity
As discussed in Section~\ref{sec2.2.1}, the PKF dynamics relies on the second-order 
flucutation-mean interaction dynamics where each prognostic function is replaced by 
a stochastic counter-part. Hence, the constructor of \texttt{SymbolicPKF} 
converts each prognostic functions as a function of an additional 
coordinate, $\omega\in\Omega$. For the Burgers' equation, $u(t,x)$ becomes $u(t,x,\omega)$.

% intro to Expectation operator
Since the computation of the second-order fluctuation-mean interaction dynamics relies 
on the expectation operator, an implementation of this expectation operator has been 
introduced in \texttt{sympkf}: it is defined as the class \texttt{Expectation} build 
by inheritance from the class \texttt{sympy.Function} so to leverage on the computational 
facilities of \texttt{sympy}. The implementation of the class \texttt{Expectation}
is based on the linearity of the mathematical expectation operator with respect 
to deterministic quantities, and its commutativity
with partial derivatives and integrals with respect to coordinates different from $\omega$ 
\eg for the Burgers' equation $\E{\partial_x u(t,x,\omega)}=\partial_x\E{u(t,x,\omega)}$. 
Note that $\E{u(t,x,\omega)}$ is a function of $(t,x)$ only: the expectation operator 
converts a random variable into a deterministic variable.

Then, the symbolic computation of the second-order fluctuation-mean interaction dynamics 
\Eq{eq_pkf_aa} is performed, thanks to \texttt{sympy}, by following the steps as described in 
Section~\ref{sec2.2.1}. In particular, the computation also leads to the tangent-linear 
dynamics of the error \Eq{eq_tl}, from which it is possible to compute the dynamics of 
the variance \Eq{eq_pkf_a} and of the metric tensor \Eq{eq_pkf_b} 
(or its associated aspect-tensor version).
Applying these steps, and the appropriate substitutions, is achieved in the back-office 
when calling the \texttt{in\_metric} or \texttt{in\_aspect} python's property 
of an instance of the class \texttt{SymbolicPKF}. This is shown for the Burgers' equation 
in \Fig{fig_burgers_unclosed}, where the background computation of the PKF dynamics 
leads to a list of the three coupled equations corresponding to the mean, 
the variance and the aspect tensor, similar to the system Eq.~(22) first obtained 
by \cite{Pannekoucke2018NPG}. 

Hence, from \sympkf, for the Burgers' equation, the VLATcov PKF dynamics given in aspect tensor reads as 
\begin{equation}\label{eq:burger_pkf_aspect}
	\left\{
\begin{array}{ccl}
	\partial_t u &=& \kappa \partial^2_x u - u \partial_x u - \frac{\partial_x V_u}{2}\\
	\partial_t V_u &=& - \frac{2 \kappa V_u}{s_{u,xx}} + 
										\kappa \partial^2_x V_u 
										- \frac{\kappa \left(\partial_x V_u\right)^{2}}{2 V_u} \\
										&&- u \partial_x V_u 
										- 2 V_u \partial_x u\\
	\partial_t s_{u,xx} &=& 
								2 \kappa s_{u,xx}^{2} {\mathbb E}\left(\eps_u \partial_x^4 \eps_u\right) 
								- 3 \kappa \partial^2_x s_{u,xx} \\
								&&- 2 \kappa 
								+ \frac{6 \kappa \left(\partial_x s_{u,xx}\right)^{2}}{s_{u,xx}} 
								- \frac{2 \kappa s_{u,xx} \partial^2_x V_u}{V_u} \\
								&&+ \frac{\kappa \partial_x V_u \partial_x s_{u,xx}}{V_u} 
								+ \frac{2 \kappa s_{u,xx} \left(\partial_x V_u\right)^{2}}{V_u^{2}} \\
								&&- u \partial_x s_{u,xx} + 2 s_{u,xx} \partial_x u	
\end{array}
\right.,
\end{equation}
where here $s_{u,xx}$ is the single component of the aspect tensor $\bs_u$ in 1D domains.
Note that in the output of the PKF equations, as reproduced in \Eq{eq:burger_pkf_aspect}, 
the expectation in the dynamics of the mean is replaced by the 
prognostic field, that is for the Burgers' equation: $\E{u}(t,x)$ is simply denoted by $u(t,x)$.

While the Burgers' equation only contains two physical processes \ie the 
non-linear advection and the diffusion, the resulting PKF dynamics 
\Eq{eq:burger_pkf_aspect} makes appear numerous terms, which justifies the use of 
symbolic computation, as above mentioned. 
The computation of the PKF dynamics 
leading to the metric and to the aspect tensor formulation takes about $1s$ of 
computation (Intel Core i7-7820HQ CPU at 2.90GHz x 8), while it can 
take more than one hour by hand.

In this example, the splitting strategy has not been considered to simplify 
the computation of the PKF dynamics. However, it can be done by considering the 
PKF dynamics for the advection $\pdt u = -u\pdx u$ and the diffusion 
$\pdt u = \kappa \pdx^2 u$, and computed separately, then merged to find the PKF dynamics
of the full Burgers' equation. For instance, the \Fig{fig_burgers_splitting} show the 
PKF dynamics for the advection (first cell) and for the diffusion (second cell), 
where the output can be trace back in \Eq{fig_burgers_unclosed} \eg 
by the terms in $\kappa$ for the diffusion.

Thanks to the symbolic computation using the expectation operator, as implemented 
by the class \texttt{Expectation}, it is possible to handle terms as 
$\E{\eps_u \partial_x^4 \eps_u}$ during the computation of the PKF dynamics. 
The next section details how these terms are handeled during the computation and 
the closure issue they bring.

\subsection{Comments on the computation of the VLATcov PKF dynamics and the closure issue}

% dire pourquoi on ne trouve pas le système d'équation Eq.22 de P18, du fait 
% de la transformation qui est réalisée pour passer d'une forme d'espérance à une 
% autre plus proche de la définition d'une fonction de corrélation... 

\subsubsection{Computation of terms $\E{\partial^\alpha \eps \partial^\beta\eps}$ and their connection to the 
correlation function}

An important point is that terms as $\E{\eps\partial^\alpha\eps}$,
\eg $\E{\eps_u\partial_x^4\eps_u}$ in \Eq{eq:burger_pkf_aspect},
are directly connected to the correlation function $\rho(\px,\py)=\E{\eps(\px)\eps(\py)}$ whose Taylor expansion is written as
\begin{equation}\label{eq:taylor}
	\rho(\px,\px+\delta\px) = \sum_k \frac{1}{k!}\E{\eps(\px)\partial^k \eps(\px)}\delta \px^k.
\end{equation}
However, during its computation, the VLATcov PKF dynamics makes appear terms 
$\E{\partial^\alpha \eps \partial^\beta\eps}$ with
$|\alpha|\leq|\beta|$,
where  for any $\alpha$, $\partial^\alpha$ 
denotes the derivative with respect to the multi-index 
$\alpha=(\alpha_{i})_{i\in[1,n]}$, $\alpha_i$ denoting the derivative order with respect to 
the $i^{th}$ coordinate $x_i$ of the coordinate system ; and 
where the sum of all derivative order is denoted by $|\alpha|=\sum_i\alpha_{i}$.
The issue it that these terms in $\E{\partial^\alpha \eps \partial^\beta\eps}$
are not directly connected to the Taylor expansion \Eq{eq:taylor}.

\begin{figure}
	\includegraphics[width=8cm]{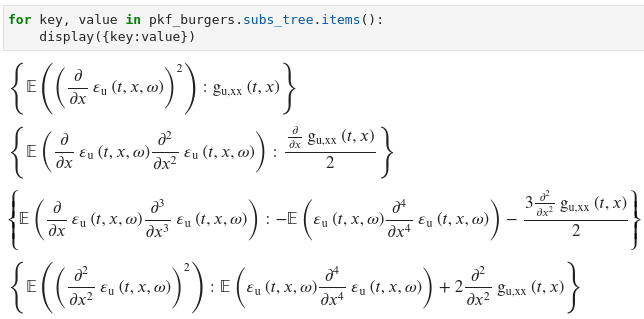}
	\caption{Substitution dictionnary computed in
	\code{sympkf} to replace terms as $\E{\partial^\alpha \eps \partial^\beta\eps}$
	by terms in $\E{\eps \partial^\gamma\eps}$ where $|\gamma|<|\alpha|+|\beta|$.
	}
	\label{fig_burgers_substree}
\end{figure}

The interesting property of these terms is that they can be rewords as
spatial derivatives of terms in the form $\E{\eps \partial^\gamma\eps}$. 
More precisely, any term $\E{\partial^\alpha \eps \partial^\beta\eps}$ can be written 
from derivative of terms in $\E{\eps \partial^\gamma\eps}$ where $|\gamma|<|\alpha|+|\beta|$,
and the term $\E{\eps \partial^{\alpha+\beta}\eps}$ (see Appendix~\ref{secB} for 
the proof). 
So, to replace any term in $\E{\partial^\alpha \eps \partial^\beta\eps}$ by 
terms in $\E{\eps \partial^\gamma\eps}$ where $|\gamma|<|\alpha|+|\beta|$,
a substitution dictionnary is computed in \sympkf, and stored as the variable 
\texttt{subs\_tree}. The computation of this substitution dictionnary is performed 
thanks to a dynamical programming strategy. Latter, the integer $|\alpha|+|\beta|$
is called the order of the term $\E{\partial^\alpha \eps \partial^\beta\eps}$.
\Fig{fig_burgers_substree} shows the substitution dictionnary computed for 
the Burgers' equation. It appears that terms of order lower than $3$ can be explicitly 
written from the metric (or its derivatives) while terms of order larger than $4$ cannot: this is known as the closure issue \citep{Pannekoucke2018NPG}.

The term $\E{\eps \partial_x^4\eps}$, which features long-range correlations, 
cannot be related neither to the variance or the metric, and 
has to be closed. We detail this point in the next section.

\subsubsection{Analytical and data-driven closure}

A naïve closure for the PKF dynamics \Eq{eq:burger_pkf_aspect} would be to replace 
the unknown term $\E{\eps_u\pdx^4\eps_u}$ by zero. 
However, in the third equation that corresponds to the aspec tensor dynamics, 
the coefficient $-3\kappa$ of the diffusion term $\pdx^2 s_u$ being 
negative, it follows that the dynamics of $s_u$ numerically explodes 
at an exponential rate. 
Of course, because the system represent the uncertainty dynamics of the Burgers' equation
\Eq{eq_burgers} that is well-posed, the parametric dynamics should not explodes. 
Hence, the unknown term $\E{\eps_u\pdx^4\eps_u}$ is crucial:
it can balance the negative diffusion so to stabilize the parametric dynamics.

For the Burgers' equation, a closure for $\E{\eps_u\pdx^4\eps_u}$ has been previously proposed \citep{Pannekoucke2018NPG}, given by 
\begin{equation}\label{eq:closureP18}
	\E{\eps_u\pdx^4\eps_u} \sim
	\frac{2}{s_u^2}\pdx^2 s_u + \frac{3}{s_u^2} - 4\frac{\left(\pdx s_u\right)^2}{s_u^3},
\end{equation}
where the symbols $\sim$ is used to indicate that this is not an equality but a proposal 
of closure for the term in the left hand side,
and which leads to the closed system
\begin{equation}\label{eq:closed_pkf}
	\left\{
	\begin{array}{ccl}	  
	  	\partial_t u &=& - u\partial_x u + \kappa\partial_x^2 u-\frac{1}{2}\partial_x V,\\
	  
	   	\partial_t V_u &=& - u\partial_x V  -2(\partial_x u) V +\kappa \partial_x^2 V \\
								   &&- \frac{\kappa}{2}\frac{1}{V}(\partial_x V)^2
								   -\frac{2\kappa}{s_{u,xx}} V_u,\\
	
		\partial_t s_{u,xx} &=&  -u\partial_x s_{u,xx}  +
				2(\partial_x u) s_{u,xx}+4\kappa\\
				&&- 2\frac{\kappa s_{u,xx}}{V_u}\partial_x^2V_u 
				 +2\frac{\kappa s_{u,xx}}{V_u^2}(\partial_x V_u)^2 \\
				&&	+\kappa\frac{1}{V_u}\partial_x V_u\partial_x s_{u,xx}
				+ \kappa \partial_x^2 s_{u,xx}\\
				&& -2\kappa\frac{1}{s_{u,xx}}(\partial_x s_{u,xx})^2.
	\end{array}
	\right.
\end{equation}
The closure \Eq{eq:closureP18} results from a local Gaussian approximation of the 
correlation function. Previous numerical experiments have shown that this closure 
is well adapted to the Burgers' equation \citep{Pannekoucke2018NPG}. 
But the approach that has been followed to find this closure 
is quite specific, and it would be interesting to design a general way to find such
a closure.

\begin{figure}
	\includegraphics[width=8cm]{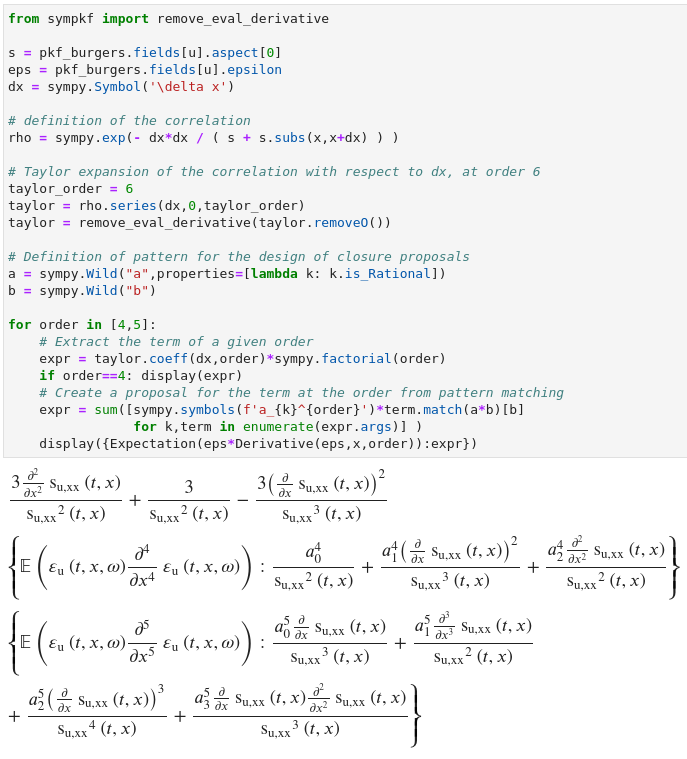}
	\caption{
		Example of a symbolic computation leading to a proposal for the closure 
		of the unknown terms of order $4$ and $5$.
	}
	\label{fig_burgers_quasi_gaussian_closure}
\end{figure}

In particular, it would be interesting to search a generic way for designing closures that leverages on the symbolic computation, which could be plugged with the PKF dynamics  
computed from \sympkf at a symbolic level. To do so, we propose an
empirical closure
which leverages on a data-driven strategy so to hybride the machine learning 
with the physics, as proposed by \cite{Pannekoucke2020GMD} with their neural network 
generator \texttt{PDE-NetGen}.

The construction of the proposal relies on the symbolic computation shown in 
\Fig{fig_burgers_quasi_gaussian_closure}:
  
The first step is to consider an analytical approximation for the correlation function.
For the illustration, we consider that the local correlation function
is well approximated by the quasi-Gaussian function
\begin{equation}\label{eq:quasigauss}
	\rho(x,x+\delta x)\approx \exp\left(- \frac{\delta x^2}{s_u(x)+s_u(x+\delta x)}\right).
\end{equation}
Then, the second step is to perform the computation of the Taylor's expansion 
of \Eq{eq:taylor} at a symbolic level. This is done thanks to \texttt{sympy}
with the method \texttt{series} applied to \Eq{eq:quasigauss} for 
$\delta x$ near the value $0$ and at a given order, \eg for the illustration
expansion is computed as the sixth order in \Fig{fig_burgers_quasi_gaussian_closure}.

Then, the identification with the Taylor's expansion \Eq{eq:taylor}, 
leads to the closure
\begin{equation}\label{eq:quasiclosure}
	\E{\eps_u\pdx^4\eps_u}\sim
	\frac{3}{s_{u,xx}^2}\pdx^2 s_{u,xx} + \frac{3}{s_{u,xx}^2} - 
	3\frac{\left(\pdx s_{u,xx}\right)^2}{s_{u,xx}^3}.
\end{equation}
While it looks like the closure \Eq{eq:closureP18}, the coefficient are not 
the same. But this suggests that the closure of $\E{\eps_u\pdx^4\eps_u}$ can be expanded  as 
\begin{equation}\label{eq:nnclosure}
	\E{\eps_u\pdx^4\eps_u}\sim
	\frac{a_0^4}{s_{u,xx}^2}\pdx^2 s_{u,xx} + \frac{a_1^4}{s_{u,xx}^2} +
	a_3^4\frac{\left(\pdx s_{u,xx}\right)^2}{s_{u,xx}^3},
\end{equation}
where $\mathbf{a}^4=(a_0^4,a_1^4,a_2^4)$ are three unknown reals. A data-driven strategy can be considered 
to find an appropriate value of $\mathbf{a}^4$ from experiments. This has been 
investigated by using the automatic generator of neural network \texttt{PDE-NetGen}
which bridges the gap between the physics and the machine-learning 
\citep{Pannekoucke2020GMD}, and where the training has lead to the value 
$\mathbf{a}^4\approx ( 0.93 ,  0.75, -1.80)
	\pm(5.1\,\, 10^{-5}, 3.6\,\, 10^{-4}, 2.7\,\, 10^{-4})$ 
(estimation obtained from 10 runs). Since this proposal is deduced from 
symbolic computation, it is easy to build some proposals for higher-order unknown terms
as it is shown in \Fig{fig_burgers_quasi_gaussian_closure} for the term 
$\E{\eps_u\pdx^5\eps_u}$.

Whatever if the closure has been obtained from an analytical or an empirical way,
it remains to compute the closed PKF dynamics to assess its performance. To 
do so a numerical implementation of the system of partial 
differential equation has to be introduced. 
As for the computation of the PKF dynamics, the design of a numerical code 
can be tedious, with a risk to introduce errors in the implementation due to 
the numerous terms occurring in the PKF dynamics.
This task can be tedious especially in reasearch when 
To facilitate the research on the PKF, \sympkf comes with a python numerical 
code generator, which provides an end-to-end investigation of the PKF dynamics.
This code generator is now detailed.

\subsection{Automatic code generation for numerical simulations}

\begin{figure}
	\includegraphics[width=8cm]{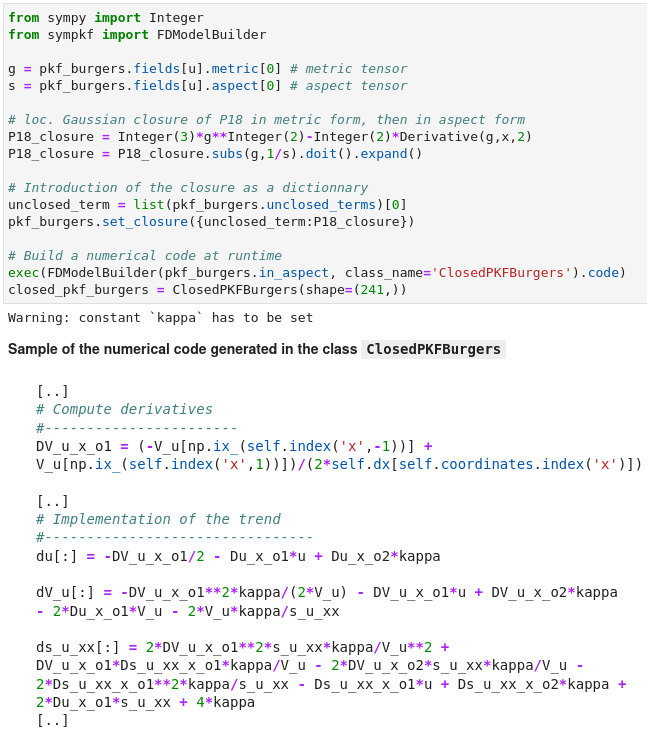}
	\caption{Introduction of a closure and automatic generation of 
	a numerical code in \sympkf.}
	\label{fig_burgers_code}
\end{figure}

\begin{figure}
	\includegraphics[width=8cm]{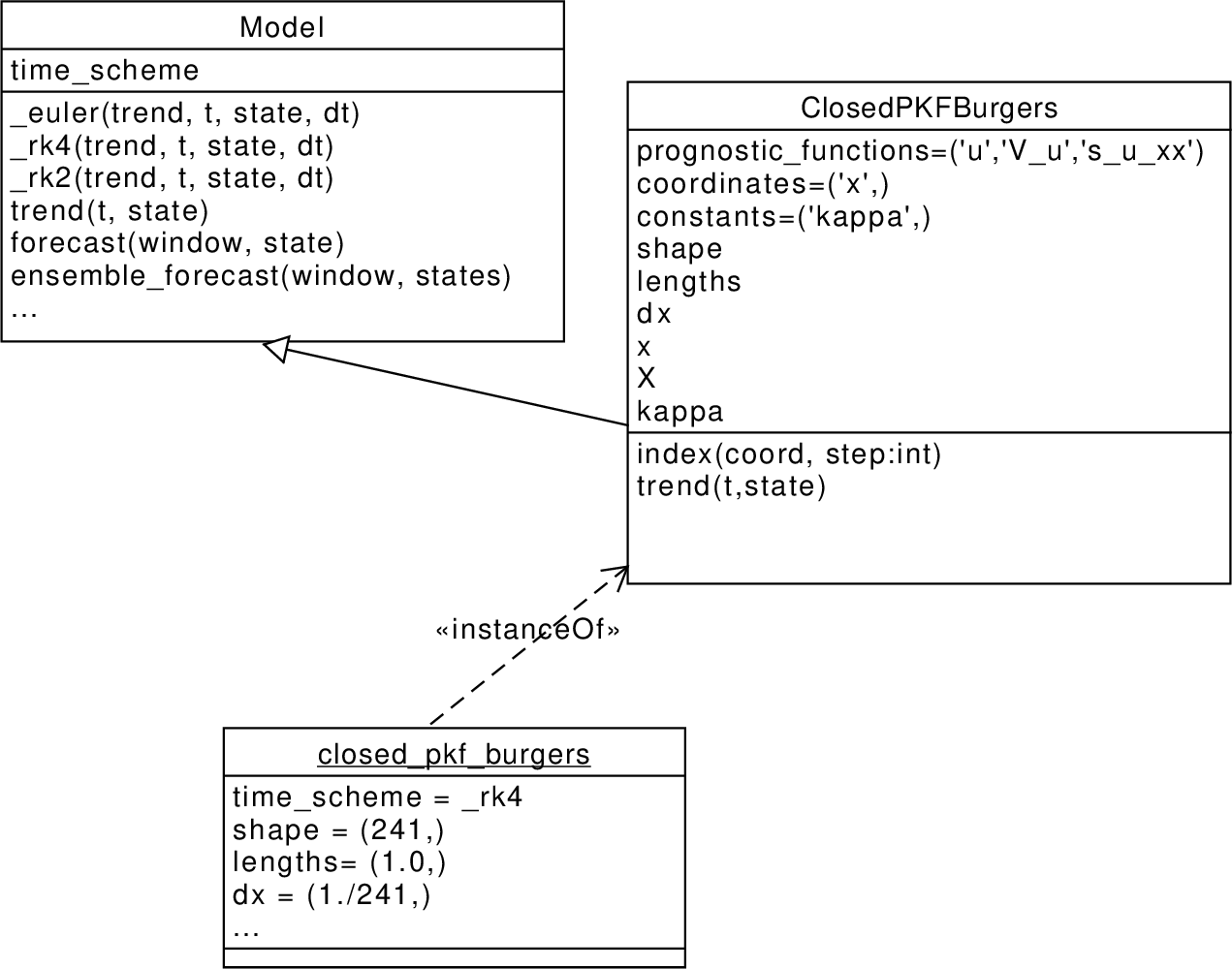}
	\caption{UML diagram showing the inheritance mechanism implemented in 
	\sympkf: the class 
	\texttt{ClosedPKFBurgers} inherits from the class \texttt{Model} 
	which implements several time schemes. Here, \texttt{closed\_pkf\_burgers}
	is an instance of the class \texttt{ClosedPKFBurgers}.
	}
	\label{fig_burgers_pkf_uml}
\end{figure}

\begin{figure*}
	\includegraphics[width=15cm]{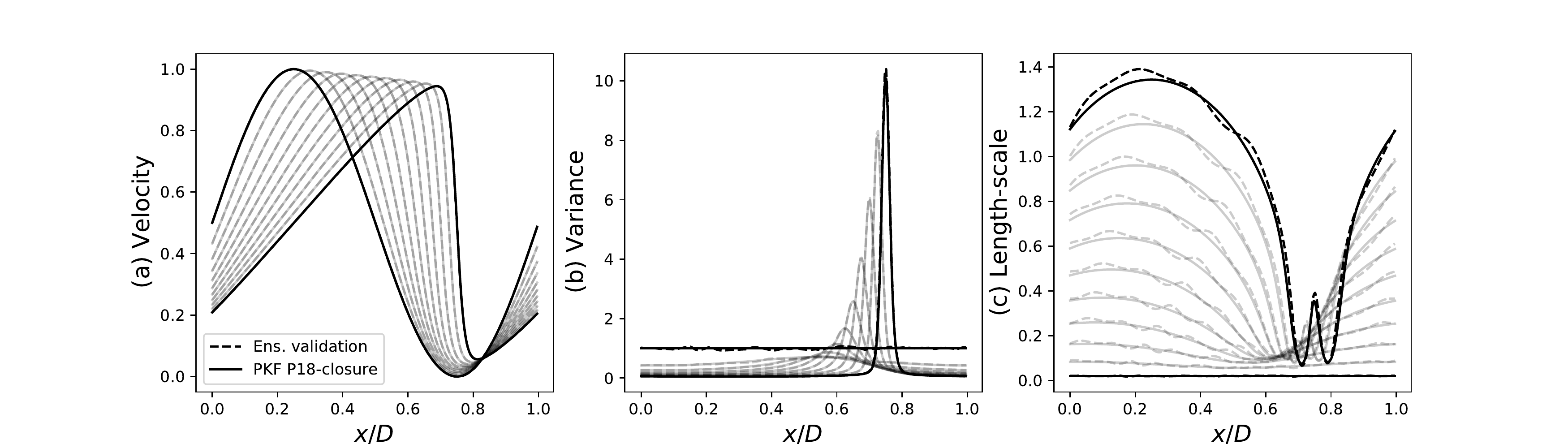}
	\caption{Illustration of a numerical simulation of the PKF dynamics 
	\Eq{eq:closureP18} (solid line), with the mean (panel a), the variance (panel b)
	and the correlation length-scale (panel c) which is defined, 
	from the component $s_{u,xx}$ of the aspect tensor, by $L(x)=\sqrt{s_{u,xx}(x)}$.
	An ensemble-based validation of the PKF dynamics is shown in dashed line.
	}
	\label{fig_burgers_pkf_simulation}
\end{figure*}

While compiled language with appropriate optimization should be important for
industrial applications, we chose to implement a pure python code generator
which offers a simple research framework for exploring the design of PKF
dynamics. 
It would have been possible code generator already based 
on \texttt{sympy} (see \eg \cite{Louboutin2019GMD}) but such code generators being 
domain specific, it was less adapted to the investigation of the PKF for arbitrary
dynamics.
In place, we consider a finite difference implementation of partial derivatives 
with respect to spatial coordinates. The default domain to perform the
computation is the periodic unit square of dimension the number of spatial coordinates.
The length of the domain can be specified along each direction.
The domain is regularly discretized along each direction while the number of 
grid-points can be specified for each direction.

The finite difference takes the form of an operator $\Fr$ that approximates any 
partial derivate at a second order of consistency:
for any multi-index $\alpha$, 
$\Fr^\alpha u \underset{0}{=}\partial^\alpha u + \Or(|\delta \textbf{x}|^2)$
where $\Or$ is the Landau's big O notation: for any $f$, the notation   
$f(\delta x) \underset{0}{=} \Or(\delta x^2)$ means that 
$\lim_{\delta x\rightarrow 0} \frac{f(\delta x)}{\delta x^2}$ is finite.
The operator $\Fr$ computed with respect to independent coordinates commute, 
\eg $\Fr_{xy} = \Fr_x\circ \Fr_y = \Fr_y\circ\Fr_x$ where $\circ$ denotes the 
composition; but it does not commute for dependent coordinates \eg
$\Fr_x^2 \neq \Fr_x\circ \Fr_x$. The finite difference of partial derivative with 
respect to multi-index is computed sequentially \eg 
$\Fr_{xxy} = \Fr^2_x\circ \Fr_y=\Fr_y\circ \Fr^2_x$. The finite difference 
of order $\alpha$ with respect to a single spatial coordinate is the centered 
finite difference based on $\alpha+1$ points.

For instance, \Fig{fig_burgers_code} shows how to close the PKF dynamics for 
the Burgers' equation following P18, and how to build a 
code from an instance of the class \texttt{sympkf.FDModelBuilder}: it creates the class 
\texttt{ClosedPKFBurgers}.
In this example, the code is executed at run time,
but it can also be written in an appropriate python's module for adapting the 
code to a particular situation or to check the correctness of the generated code.
At the end, the instance \texttt{closed\_pkf\_burgers} of the class 
\texttt{ClosedPKFBurgers} is created, raising a warning to indicate that the 
value of constant $\kappa$ has to be specified before to perform a numerical simulation.
Note that it is possible to set the value of kappa as a keyword argument in the 
class \texttt{ClosedPKFBurgers}.
\Fig{fig_burgers_code} also show a sample of the generated code 
with the implementation of the computation of the first order partial derivative 
$\pdx V_u$, which appears as a centered finite difference.
Then, the sample of code show how the partial derivatives are used to compute the 
trend of the system of partial differential equations \Eq{eq:closed_pkf}.

The numerical integration is handle through the inheritance 
mechanism: the class \texttt{ClosedPKFBurgers} inherits the integration 
time loop from the class \texttt{sympkf.Model} as described by the 
UML diagram shown in \Fig{fig_burgers_pkf_uml}.
In particular, the class \texttt{Model} contains several time schemes \eg a fourth-order 
Runge-Kutta scheme. The details of the instance \texttt{closed\_pkf\_burgers} 
of the class \texttt{ClosedPKFBurgers} make appears that 
the closed system \Eq{eq:closed_pkf} will be integrated by using a RK4 time scheme,
on the segment $[0,D]$ (here $D=1$) with periodic boundaries, and discretized by $241$
points.

Thanks to the end-to-end framework proposed in \sympkf, it is possible to perform
a numerical simulation based on the PKF dynamics \Eq{eq:closureP18}.
To do so, we set $\kappa= 0.0025$ and consider the simulation starting from the 
Gaussian distribution $\mathcal{N}(u_0,\mP^f_h)$ of mean 
$u_0(x) = U_{max}[1+\cos(2\pi(x-D/4)/D)]/2$ with $U_{max}=0.5$, and 
of covariance matrix
\begin{equation}\label{eq:rhoh}
	\mP^f_h(x,y) = V_h \exp\left(- \frac{(x-y)^2}{2l_h^2}\right),
\end{equation}
where $V_h = 0.01 U_{max}$ and $l_h=0.02D\approx 5 dx$. The time step of the fourth-order 
Runge-Kutta scheme is $dt=0.002$.
The evolution predicted from the PKF is shown in 
\Fig{fig_burgers_pkf_simulation} (solid lines). This simulation illustrates 
the time evolution of the mean (panel a) and of the variance (panel b) ; 
the panel (c) represents the evolution of the correlation length-scale defined 
from the aspect tensor as $L(x)=\sqrt{s_{u,xx}(x)}$. Note that at time $0$, 
the length-scale field is $L(x)=l_h$. 
For the illustrations, the variance (the length-scale) is normalized by its initial 
value $V_h$ ($l_h$).

In order to show the skill of the PKF applied on the 
Burgers, when using the closure of P18, an ensemble validation is now performed.
Note that the code generator of \sympkf can be used for an arbitrary dynamics \eg 
the Burgers' equation itself. Hence, a numerical code solving the Burgers' equation is 
rendered from its symbolic definition. Then an ensemble of $1600$ forecasts is computed 
starting from an ensemble of initial errors at time $0$. The ensemble of initial 
errors is sampled from the Gaussian distribution $\mathcal{N}\left(0,\mP^f_h\right)$
of zero mean and covariance matrix $\mP^f_h$.
Note that the ensemble forecasting implemented in \sympkf as the method 
\texttt{Model.ensemble\_forecast} (see \Fig{fig_burgers_pkf_uml}) 
leverages on the multiprocessing tools of python, 
so to use the multiple cores of the CPU, when present. On the computer used for the 
simulation, the forecasts are performed in parallel on the $8$ cores.
The ensemble estimation of the mean, the variance and the length-scale are shown in 
\Fig{fig_burgers_pkf_simulation} (dashed lines). Since the ensemble is finite, a sampling 
noise is visible \eg on the variance at the initial time that is not strictly equals 
to $V_h$. In this simulation, it appears that 
the PKF (solid line) coincide with the ensemble estimation (dashed lines) which shows 
the ability of the PKF to predict the forecast-error covariance dynamics.
Note that the notebook corresponding to the Burgers' experiment is available in 
the example directory of \sympkf.

While this example shows an illustration of \sympkf in 1D domain, 
the package also applies in 2D and 3D domains, as presented now.

\subsection{Illustration of a dynamics in a 2D domain}\label{Sec4.5}

\begin{figure*}
	\includegraphics[width=12cm]{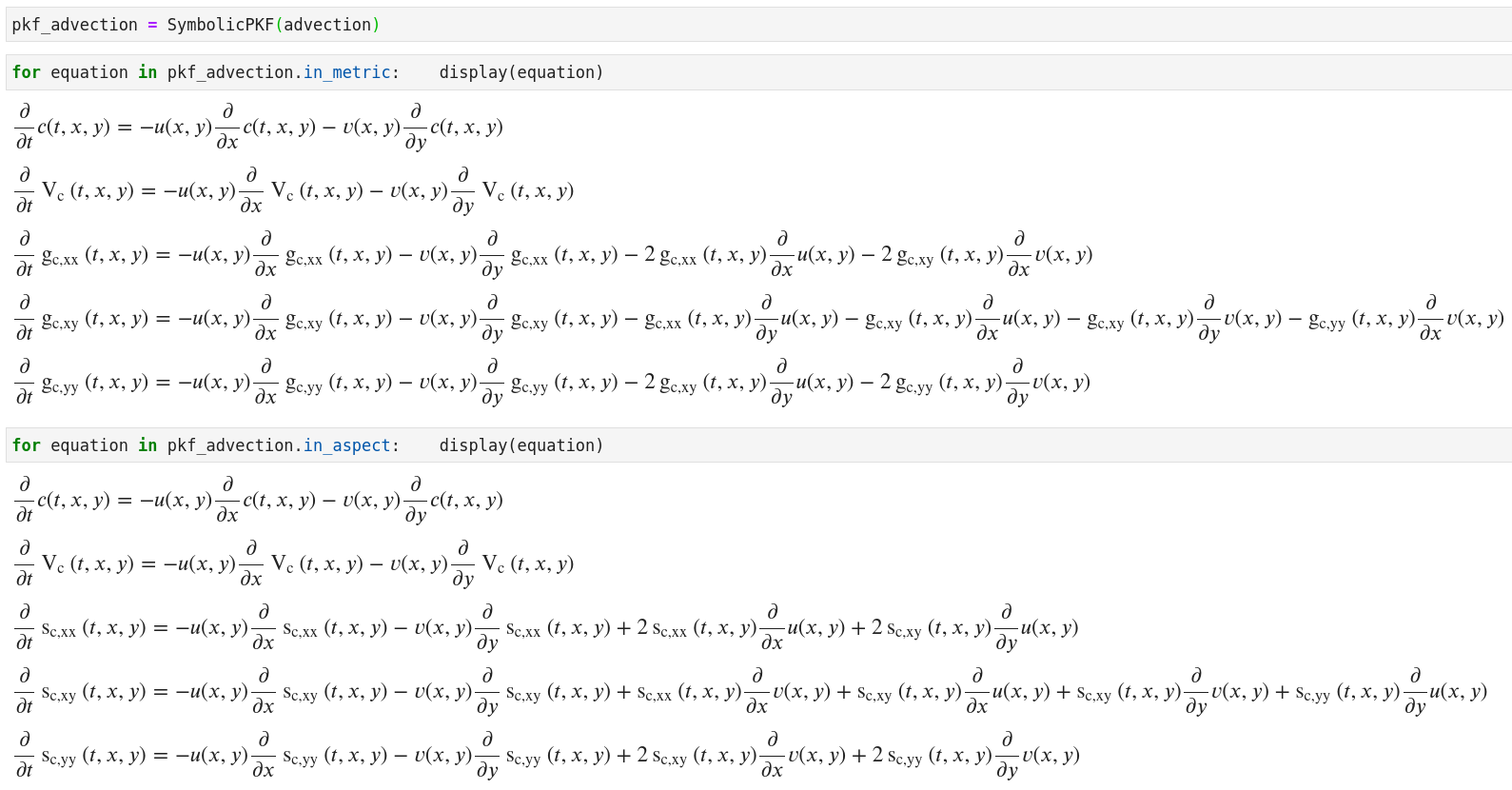}
	\caption{Sample of code and Jupyter notebook outputs: 
	system of partial differential equations produced by \code{sympkf}
	when applied to the linear advection \Eq{eq_linearAdvection}.}
	\label{fig_advection_pkf}
\end{figure*}

In order to illustrate the ability of \sympkf to apply in a 2D or a 3D domain, we 
consider the linear advection of a scalar field $c(t,x,y)$ by a stationary velocity 
field $\bu=(u(x,y),v(x,y))$, which reads as the partial differential equation
\begin{equation}\label{eq_linearAdvection}
	\pdt c + \mathbf{u} \nabla c =0.
\end{equation}
As for the Burgers' equation, the definition of the dynamics relies on sympy 
(not shown but similar to the definition of the Burgers' equation as given in 
\Fig{fig_burgers_setup}).
%as shown in \Fig{fig_advection_setup}. 
This leads to preprocess the dynamics by creating the instance \code{advection}  
of the class \code{PDESystem},
which transform the equation into a system of partial differential equations.  
In particular, the procedure will diagnose the prognostic functions of a dynamics, 
here the function $c$ ; the constant functions, here $u,v$ the component of the 
velocity $(u,v)$ ; for this example there is no constant nor exogenous function.

The calculation of the parametric dynamics is handled by the class \code{SymbolicPKF} 
as shown in the first cell in \Fig{fig_advection_pkf}. 
The parametric dynamics is a property of the instance \code{pkf\_advection} of the 
class \code{SymbolicPKF}, and when it is called, the parametric dynamics is computed 
once time.
The parametric dynamics formulated in terms of metric is first computed, 
see the second cell.
For the 2D linear advection, the parametric dynamics is a system of 
five partial differential equations, as it is shown in the output of the second cell:
the dynamics of the ensemble average $\E{c}$ which outputs as $c$ for the 
sake of simplicity (first equation), the dynamics of the variance 
(second equation) and the dynamics of the local metric tensor (last three equations). 
In compact form, the dynamics is given by the system
\begin{subequations}\label{eq_apm}
\begin{align}
	\partial_t c +\bu\nabla c &= 0,\label{eq_apm_1} \\
	\partial_t V_c +\bu\nabla V_c &= 0,\label{eq_apm_2} \\
	\partial_t \bg_c +\bu\nabla\bg_c &= -\bg_c\left(\nabla\bu\right) - 
		\left(\nabla\bu\right)^T\bg_c, \label{eq_apm_3}
\end{align}
\end{subequations}
which corresponds to the 2D extension of the 1D dynamics 
first found by \cite{Cohn1993MWR} \citep{Pannekoucke2016T}, 
and validates the computation performed in \sympkf.
Due to the linearity of the linear advection \Eq{eq_linearAdvection}, 
the ensemble average \Eq{eq_apm_1} is governed by the same dynamics 
\Eq{eq_linearAdvection}. The variance is advected by the flow While both the variance, 
\Eq{eq_apm_2}, and the metric are advected by the flow, the metric is also deformed by the 
shear \Eq{eq_apm_3}. 
This deformation more commonly appears on the dynamics written in aspect tensor form, 
which is given by 
\begin{subequations}\label{eq_apd}
	\begin{align}
		\partial_t c +\bu\nabla c &= 0,\\
		\partial_t V_c +\bu\nabla V_c &= 0,\\
		\partial_t \bs_c +\bu\nabla\bs_c &= \left(\nabla\bu\right) \bs_c+ \bs_c\left(\nabla\bu\right)^T,\label{eq_apdc}
	\end{align}
\end{subequations}
where \Eq{eq_apdc} is similar to the dynamics of the conformation
tensor in viscoelastic flow \citep{Bird1995FM,Hameduddin2018JFM}.

This example illustrates a 2D situation, but it runs as well in 3D too.
Similarly to the simulation conducted for the Burgers' equation, it is possible to 
automatically generate a numerical able to perform numerical simulations of the dynamics
\Eq{eq_apd} (not shown here).
Hence, this 2D domain example showed the ability of \sympkf to apply in dimensions 
lager than the 1D. 

Before to conclude, we would like to present a preliminary application of \sympkf 
in a multivariate situation.

\subsection{Toward the PKF for multivariate dynamics}

\sympkf  can be used to compute the prediction of the variance and the anisotropy 
in a multivariate situation. 

Note that one of the difficulty with the multivariate situation is that the number 
of equations increases linearly with the number of fields and the dimension of 
the domain \eg for a 1D (2D) domain and two multivariate physical fields, 
there are two ensemble averaged fields, two variance fields and two (six) 
metric fields. Of course this is no not a problem when using a computer 
algebra system as done in \sympkf.

\begin{figure}
	\includegraphics[width=8cm]{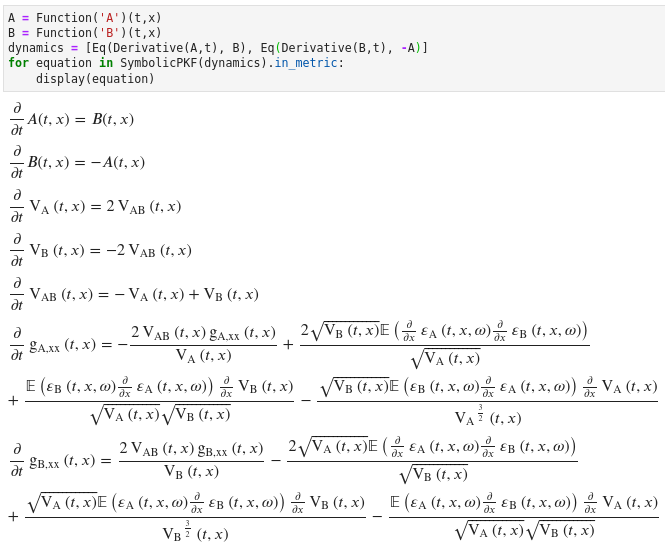}
	\caption{Illustration of the computation of the PKF dynamics for 
	a simple multivariate situation by using \sympkf.
	}
	\label{fig_multivariate}
\end{figure}

So to illustrate the multivariate situation, only a very simple example is introduced.
Inspired from chemical transport models encountered in air quality, we 
consider the transport, over a 1D domain, of two chemical species, whose the concentrations 
are denoted by $A(t,x)$ and $(B(t,x)$,  and advected by the wind $u(x)$. 
For the sake of simplicity, the two species interact following a periodic dynamics, 
leading to the coupled system 
\begin{subequations}\label{eq:ctm1D}
	\begin{eqnarray}
		\partial_t A + u\partial_x A &=& B,\\
		\partial_t B + u\partial_x B &=& -A,
	\end{eqnarray}
\end{subequations}
Thanks to the splitting strategy, the PKF dynamics due to the advection 
has already be detailed in the previous section (see Section~\ref{Sec4.5}), 
so we can focus on the chemical part of the dynamics which is given by the 
processes on the right hand side of \Eq{eq:ctm1D}.
The PKF of the chemical part is computed thanks to \sympkf, and shown 
in \Fig{fig_multivariate}.
This time, and as it is expected, multivariate statistics appear in the dynamics.
Here, the dynamics of the cross-covariance $V_{AB}=\E{e_A e_B}$
is given by the fifth equation.
The coupling makes appear unknown terms \eg the term 
$\E{\partial_x\eps_A\partial_x\eps_B}$ which appears in the sixth equation.
To go further, some research is still needed to explore the 
dynamics and the modelling of the multivariate cross covariances. 
A possible 
direction is to take advantage of the multivariate covariance model based on 
balance operator as often introduced in variational data assimilation 
\citep{Derber1999TA,Ricci2005MWR}. Note that such multivariate covariance models 
has been recently considered for the design of the multivariate PKF analysis step 
\citep{Pannekoucke2021TA}. Another way is to consider a data-driven stragy to learn the 
physics of the unknown terms from a training based on ensembles of forecasts 
\citep{Pannekoucke2020GMD}.

To conclude, this example shows the potential of interest of \sympkf 
to tackle the multivariate situation. Moreover, the example also shows that 
\sympkf is able to perform the PKF computation for a system of partial differential
equations. However all the equations should be prognostic, \sympkf is not 
able to handle diagnostic equations.

\section{Conclusion}\label{sec5}

This contribution introduced the package \sympkf that can be used to conduct the 
research on the parametric Kalman filter prediction step for
covariance models parameterized by the variance and the anisotropy (VLATcov), 
by providing an end-to-end framework: from the equations of 
a dynamics to the development of a numerical code. 

The package has been first introduced by considering the
non-linear diffusive advection dynamics, the Burgers' equation. In particular 
this example shown the ability of \sympkf to handle abstract terms \eg
the unclosed terms formulated with the expectation operator. The expectation 
operator implemented in \sympkf is a key tool for the computation of the 
PKF dynamics. Moreover, we showed how to handle closure and how to automatically 
render numerical codes.

For univariate situations, \sympkf applies in 1D domain as well as in 
2D and 3D domains. This has been shown by considering the computation of the 
PKF dynamics for the linear advection equation on a 2D domain. 

A preliminary illustration on 
a multivariate dynamics showed the potential of \sympkf to handle the 
dynamics of multivariate covariance. But this point has to be further 
investigated, and this constitutes the main perspective of development. 
Moreover, to perform a multivariate assimilation cycle with the PKF, the 
multivariate formulation of the PKF analysis state is needed. A first investigation 
of the multivariate PKF has been proposed by \cite{Pannekoucke2021TA}.

In its present implementation, \sympkf is limited to the computation with 
prognostic equations. It is not possible to consider dynamics based on diagnostic 
equations while these are often encountered in atmospheric fluid dynamics \eg 
the geostrophic balance. This constitutes another topic of research development
for the PKF, facilitated by the use of symbolic exploration.

Note that the expectation operator as introduced here can be used to compute 
Reynolds equations encountered in turbulence. This open new perspectives
of use of \sympkf for other applications that could be interesting especially
for automatic code generation.

%%%%%%%%%%%%%%%%%%%%%%%%%%%%%%%%%%%%%%%%%%%%%%%%%%%%%%%%%%%%%%
\appendix
%%%%%%%%%%%%%%%%%%%%%%%%%%%%%%%%%%%%%%%%%%%%%%%%%%%%%%%%%%%%%%
%\IncludePDF{./path/paper.pdf} 

\section{Splitting for the computation of the parametric dynamics}\label{secA}

In this section we show that using a splitting strategy is possible for the design of
the parametric dynamics. For this, it is enough to show that given a dynamics written
as 
\begin{equation}\label{ap_eq_mul_proc}
	\pdt\xs = f_1(\xs)+f_2(\xs),
\end{equation}
the dynamics of the error, the variance, the metric and the aspect all write 
as a sum of trends depending on each processes $f_1$ and $f_2$. We show this starting 
from the dynamics of the error. 

Due to the linearity of the derivative operator, the TL dynamics resulting from \Eq{ap_eq_mul_proc}
writes 
\begin{equation}\label{ap_eq_e}
	\pdt e = f_1'(e) + f_2'(e).
\end{equation}
which can be written as the sum of two trends $\pdt e_1 = f_1'(e)$ and $\pdt e_2 = f_2'(e)$, 
depending exclusively on $f_1$ and $f_2$ respectively.
For the variance's dynamics, $\pdt V = 2\E{e\pdt e}$, substitution by \Eq{ap_eq_e} leads to 
\begin{equation}\label{ap_eq_variance}
	\pdt V = \pdt V_1 + \pdt V_2,
\end{equation}
where $\pdt V_1 = 2\E{e f_1'(e)}$ and $\pdt V_2 = 2\E{e f_2'(e)}$,
depends exclusively on $f_1$ and $f_2$ respectively.
Then the standard deviation dynamics, obtained by differenciating $\sigma^2=V$ as 
$2\sigma\pdt\sigma = \pdt V$, 
\begin{equation}
	\pdt \sigma = \frac{1}{\sigma}\pdt V_1 + \frac{1}{\sigma}\pdt V_2,
\end{equation}
writes as the sum of two trends $\pdt \sigma_1 = \frac{1}{\sigma}\pdt V_1$ and 
$\pdt \sigma_2 = \frac{1}{\sigma}\pdt V_2$, depending exclusively on $f_1$ and $f_2$ respectively.
It results that the dynamics of the normalized error $\eps = \frac{1}{\sigma}e$, deduced from the time derivative of 
$e=\sigma\eps$, $\pdt e= \eps\pdt\sigma + \sigma\pdt \eps$, writes
\begin{equation}
	\pdt\eps = 
		\frac{1}{\sigma}\left[f_1'(e) -\frac{\eps}{2\sigma}\pdt V_1 \right]+
		\frac{1}{\sigma}\left[f_2'(e) -\frac{\eps}{2\sigma}\pdt V_2 \right]
\end{equation}
and also expands as the sum of two trends $\pdt \eps_1=\frac{1}{\sigma}\left[f_1'(e) -\frac{\eps}{2\sigma}\pdt V_1 \right]$ 
and $\pdt\eps_2=\frac{1}{\sigma}\left[f_2'(e) -\frac{\eps}{2\sigma}\pdt V_2 \right]$, again depending exclusively on $f_1$ 
and $f_2$ respectively.
For the metric terms $g_{ij}=\E{\partial_i\eps\partial_j\eps}$,
we deduce that the dynamics $\pdt g_{ij} = \E{\partial_i(\pdt\eps)\partial_j\eps}+\E{\partial_i\eps\partial_j(\pdt\eps)}$
expands as 
\begin{equation}\label{ap_eq_metric}
	\pdt g_{ij} = \pdt {g_{ij}}_1 +\pdt {g_{ij}}_2,
\end{equation}
with $\pdt {g_{ij}}_1 = \E{\partial_i(\pdt\eps_1)\partial_j\eps}+\E{\partial_i\eps\partial_j(\pdt\eps_1)}$
and $\pdt {g_{ij}}_2 = \E{\partial_i(\pdt\eps_2)\partial_j\eps}+\E{\partial_i\eps\partial_j(\pdt\eps_2)}$
where each partial trend depends exclusively on $f_1$ and $f_2$ respectively.
To end, dynamics of the aspect tensor $\bs$ is deduced from \Eq{eq_in_aspect} 
which expands as 
\begin{equation}\label{ap_eq_aspect}
	\pdt \bs = \pdt\bs_1 + \pdt\bs_2,
\end{equation}
where 
$\pdt\bs_1 = -\bs(\pdt{\bg}_1)\bs$ 
and 
$\pdt\bnu_2 = -\bs(\pdt{\bg}_2)\bs$ 
only depend of on $f_1$ and $f_2$ respectively.

To conclude, the computation of the parametric dynamics for \Eq{ap_eq_mul_proc}, is deduced 
from the parametric dynamics of $\pdt\xs = f_1(\xs)$ and $\pdt\xs = f_2(\xs)$
calculated separately, then merged together to obtain the dynamics of the variance \Eq{ap_eq_variance},
of the metric \Eq{ap_eq_metric} and of the diffusion \Eq{ap_eq_aspect}

\section{Computation of terms $\E{\partial^\alpha\varepsilon\partial^\beta\varepsilon}$}\label{secB}

In this section we proof the property

\begin{property}\label{theo1}
$\E{\partial^\alpha\varepsilon\partial^\beta\varepsilon}$ 
$|\alpha|\leq|\beta|$,
can be related to the correlation 
expansion terms $\E{\eps\partial^\gamma\eps}$ 
where $|\gamma|<|\alpha|+|\beta|$, and the term $\E{\eps \partial^{\alpha+\beta}\eps}$. 
\end{property}

 The derivative with respect to a zero $\alpha_i$
is the identity operator. Note that the multi-index forms a semi-group since 
for two multi-index $\alpha$ and $\beta$ we can form the multi-index 
$\alpha+\beta = (\alpha_i+\beta_i)_{i\in[1,n]}$.

Now the property \Eq{theo1} can be proven considering the following recurrent process:

Asuming that the property is true for all patterns of degree strictly lower to
the degree $|\alpha|+|\beta|$
Without loss of generality we assume  $\alpha_i>0$ and denote $\delta_i=(\delta_{ij})_{j\in[1,n]}$ where $\delta_{ij}$ is the Kroenecker symbol 
($\delta_{ii}=1$, $\delta_{ij}=0 \text{ for } j\neq i$). From the formula
\begin{equation}
\partial_{x^i}\left(
\partial^{\alpha-\delta_i}\eps \partial^\beta\eps
\right) = 
\partial^\alpha\eps \partial^\beta\eps + 
\partial^{\alpha-\delta_i}\eps\partial^{\beta+\delta_i}\eps
\end{equation}
and from the commutativity of the expectation operator and the partial derivative with respect to the coordinate system, it results that 
\begin{equation}
\E{\partial^\alpha\eps \partial^\beta\eps} =
 \partial_{x^i}\E{\partial^{\alpha-\delta_i}\eps \partial^\beta\eps}
-\E{\partial^{\alpha-\delta_i}\eps\partial^{\beta+\delta_i}\eps}.
\end{equation}
Considereing the terms of the left hand side. In one hand, we observe that the 
degree of the first term is decreasing to $|\alpha|+|\beta|-1$, from the reccurence assumption
$\E{\partial^{\alpha-\delta_i}\eps \partial^\beta\eps}$ can be expanded as terms of the form 
$\E{\eps\partial^\gamma\eps}$. On the other hand, the degree of the 
second term remains the same, $|\alpha|+|\beta|$, but with a shift of the derivative order.
This shift of the order can be do it again following the same process, leading after iterations to 
the term $\E{\eps\partial^{\alpha+\beta}\eps}$.

%\addcontentsline{toc}{chapter}{Bibliographie}
\bibliographystyle{ametsoc}
\bibliography{paper}
%\bibliography{database}

%\printindex

\end{document}